\documentclass[12pt]{article}
\usepackage{graphicx} 
\usepackage[a4paper, total={7in, 9in}]{geometry}
\usepackage{xcolor}
\usepackage[sort]{cite}
\usepackage{amssymb}
\usepackage{hyperref}
\usepackage{rotating}
\usepackage{amsmath}
\usepackage{tcolorbox}
\usepackage{fancyhdr}

\usepackage{booktabs} 
\usepackage{wrapfig}
\usepackage{subcaption}
\usepackage{lipsum} 
\fancypagestyle{firstpage}{%
  \fancyhead{}
  
  \fancyfoot[L]{\vspace{10pt} \small{The authors would like to thank the mathematical research institute MATRIX in Australia where part of this research was performed.}}
}

\begin{document}
\title{Exploring Drivers of Extreme Housing Price in Australia\\
       \large Of \textit{price} and men: How well-intentioned regulations are hurting the economy.}
\date{}
\author{Grace Burtenshaw$^1$, Ashley Burtenshaw${^2}$, Meagan Carney$^{1*}$}
\maketitle
\thispagestyle{firstpage}
\thanks{\noindent$^*$Corresponding author email: m.carney@uq.edu.au \\$^1$The University of Queensland, School of Mathematics and Physics, Australia\\ $^2$ Barings Global Asset Based Finance | Australia}

\begin{abstract}
In recent years Australia has observed a growing, unexplained resilience of increasing house price trends. Here, we seek to understand what is driving Australia's indestructible asset using insights from market experts. We construct a differential equation model of house price to develop intuition for its historical behaviour and responsiveness to changes in mortgage rates. Using this model, we identify a point of 'decoupling' between house price and mortgage rate in the system with supply limitations found to be the main driver for this change. From there, modern extreme value techniques are implemented on real-world data to investigate how the effectiveness of mortgage rate in moderating extreme house price has changed before and after this historical decoupling. We find that without an increase in the housing supply chain, through either deregulation or reduced competition with government building, an 11\% increase in mortgage rate will be needed to slow extreme housing costs.
\end{abstract}

\section{Background and Introduction}
\subsection{The Housing Crisis in Australia}
Affordable housing is a key component for any thriving economy. It is a basic resource that shapes the well-being of individuals, communities and cities \cite{science_direct_housing_2024}. In recent years, the Australian economy has been eclipsed by a severe housing crisis. Deigned a "wicked problem", the Australian housing crisis is characterised by a rapid growth in rental prices, climbing mortgage rates, and a significant imbalance between demand and supply in the market \cite{abc_wicked_problem}. In the past 25 years, Australia has observed a doubling of house prices relative to income, accompanied with some of the lowest rates of building approvals \cite{economist_carrie_index,apu_building_approvals}.

Despite continued government intervention, most notably the implementation of Housing Australia's Home Guarantee Scheme, the Reserve Bank of Australia (RBA) has cautioned that increased interest rates and inflation are expected to continue \cite{afr_inflation_2026}. Historically, lowered mortgage rates were offered as an incentive to the market to offset rising house prices. Comparatively, increased mortgage rates have been used to decrease demand in an effort to forcefully lower house prices (or prevent them from rising). In recent years, Australian economists have anecdotally observed an ineffectiveness in the impact of this strategy on the market, illuminating an unexpected resilience in rising house prices to mortgage rate changes \cite{economist_affordability}.  Therefore, the goal of this article is to analytically investigate this observed resilience, determine the drivers that influence it, and propose changes in public policy. We achieve this using a combination of historical data analysis and robust mathematical and statistical modelling techniques.

Australian housing research has been dominated by econometric and policy orientated models, with notably fewer statistical approaches. One such vein uses traditional autoregressive conditional heteroskedasticity models (ARCH or GARCH), and correlation modelling to capture volatility \cite{qld_renting_reforms, prres_derivatives_2011, rent_control_seinfeld, semantic_scholar_rent_control}. These approaches often rely on restrictive assumptions and focus primarily on conditional variance. By implementing a blend of statistical modelling techniques, we are able to capture system dynamics, fat tails, asymmetry, and extremes, which can otherwise prove difficult to do. 
 
Where our work differs again is through the inclusion of mathematical modelling techniques that enable the study of an emerging decoupling phenomenon between mortgage rates and house prices, providing insight into the sudden resilience observed in the housing market. Although these dynamics have been recognised in a variety of economic and modelling work, papers by \cite{bonga_bonga_contagion_2025,witt_decoupling_2023,chittedi_decoupling_2020} explore these problems from a macroeconomic lens and mainly describe the relationships in observed data. In this setting, the inclusion of ordinary differential equation (ODE) models allows us to explicitly represent the dynamical system underpinning house price, whilst simultaneously exploring the relationships between supply, demand, and policy interventions. The work by Caldera and Johansson uses a similar approach to model price responsiveness in OECD countries
\cite{caldera_johansson_2013}. 

Further, there exists extensive literature on various macroeconomic indicators and drivers of market resilience in Australia. Many works investigate the drivers of the crisis by looking at housing through a systematic or embedded network system and using content analysis \cite{pablo_littleton_london_2024, ayub_urban_resilience_2020}. Although content analysis is useful for capturing the controversy and widespread conversation surrounding housing crises, statistical analysis offers a more rigorous framework--especially when exploring and quantifying the impact of underlying drivers of the market behaviour. In line with recent investigations conducted by Saunders and Tulip, we also consider a more detailed range of drivers (i.e. dwelling completion rates, approvals, net migration) \cite{saunders_tulip_2019}. By examining how these drivers and policies shift the underpinning dynamics of house price, we are able to forecast without having to make methodological assumptions common in structural macroeconometric modelling. One key example is that we do not require our data to be stationary. 

Our model addresses a gap in the literature by taking a blended ODE and extreme value analysis approach to understand how and why, drivers and dynamic shifts contribute and impact house price allowing us to target problems at their source. Using these techniques and insight, we build forecasts which aim to reduce uncertainty and inform decision making surrounding the policy reform targeting current price resilience. 

This paper is outlined in the following way: 
\begin{itemize}
\item Section \ref{data} describes the results of a preliminary data-based investigations on common drivers of house prices using dwelling value as the outcome variable.
\item Section \ref{methods} uses mathematical modelling and extreme value theory to investigate the diminishing effect of mortgage rate and consumer price index on house price.
\item Section \ref{discussion} presents our conclusions in the context of the economy and currently debated pubic policy.
\end{itemize}
\clearpage
\subsection{Data-driven Investigation of House Prices}\label{data}
We begin by investigating the relationships between dwelling value and commonly known drivers using historical data taken from the Australian Bureau of Statistics and Bloomberg. A summary of each historical time-series can be found in Table \ref{tab:ts}. In our data-based investigation, we only use historical data points with overlapping dates across all time-series.
\begin{wrapfigure}{l}{0.4\textwidth}
\centering
\includegraphics[width=\linewidth]{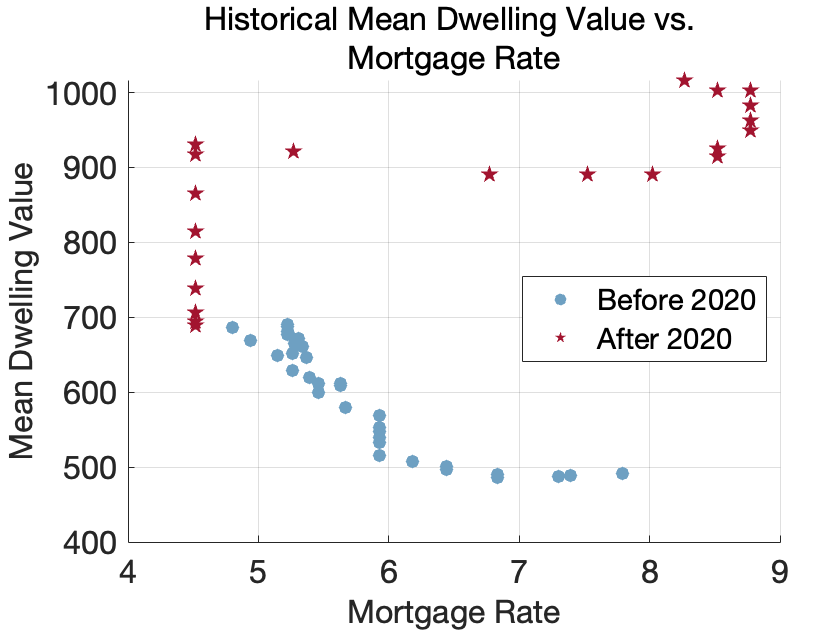}
\caption{Quarterly mortgage rate against corresponding mean dwelling value for the years 2011 - 2025}\label{fig:pricerate}
\end{wrapfigure}
Importantly, we find that mortgage rate and mean dwelling value follow a generalised logistic relationship for the years prior to 2020, after which this relationship no longer holds (Figure \ref{fig:pricerate}). Such an observation indicates that mortgage rate had a decreasing effect on dwelling values prior to 2020; however, this effect seems to be significantly diminished after mortgage rate was kept consistently low over the period 2020-2021. For the purposes of this investigation, we refer to this phenomenon as a \textit{decoupling} of the mortgage rate - mean dwelling value relationship. To understand the driving force behind this phenomenon of decoupling, we first investigate common measures of demand and supply as contributing factors: national population increase and dwelling completion value, respectively. We then use these data-based insights to build a comprehensive mathematical model to analytically understand the more complex dynamical relationships that may be contributing to this observed decoupling.

\begin{wrapfigure}{r}{0.7\textwidth}
\centering
\includegraphics[width=\linewidth]{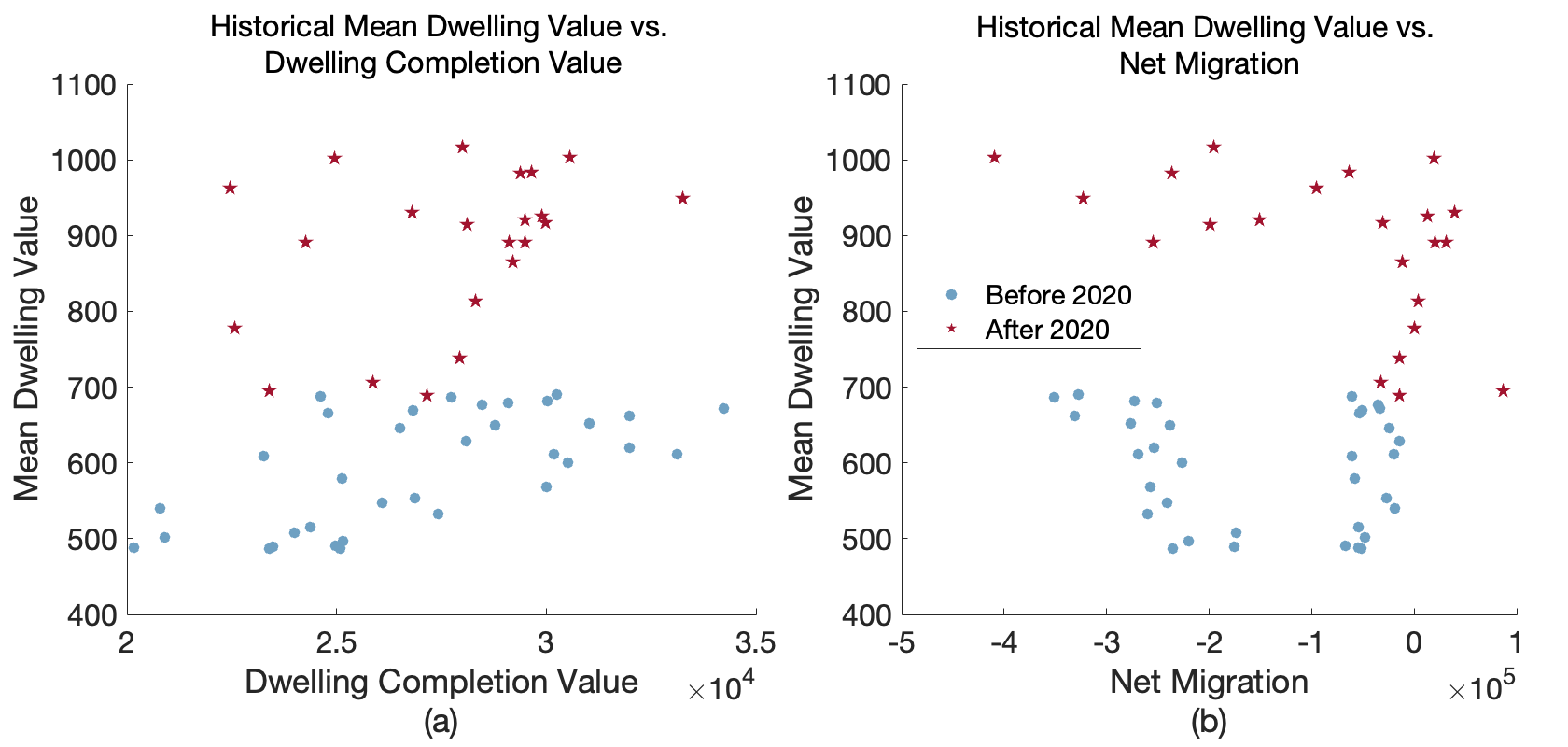}
\caption{Quarterly (a) dwelling completion values and (b) net migration rates against corresponding mean dwelling value for the years 2011 - 2025}
\end{wrapfigure}\label{fig:drivers}

The Australian Bureau of Statistics reports an increase in the total population of Australia that is largely driven by net overseas migration. Interestingly, we observe no direct relationship between net migration and mean dwelling value (Figure \ref{fig:drivers}). This holds true in the years before 2020 ($p=0.27$) and after 2020 ($p=0.06$). Next, we consider dwelling completion values as a measure of housing supply across Australia. We find that prior to 2020, dwelling completion values had a positive linear relationship to mean dwelling value ($p<10^{-5}$). No relationship is observed after 2020 ($p=0.09$). In general, this means that dwelling completions were performed at an increased rate compared to demand prior to 2020 and performed at the same rate as demand after 2020. This indicates a supply-chain problem for completions beyond already built dwellings in the years after 2020. 

Whilst no relationship between mean dwelling value and net overseas migration was observed, it is still natural to assume some underlying increase in demand for housing on the whole. Indeed, in the following section we find that demand is becoming an increasingly important driver in the observed resilience of house price. Whether this demand is driven by renter to home-owner transitions, increases in populations of ages in home-owner ranges, or increases in the number of single-owner properties across Australia is outside the scope of this investigation which is focused on the supply-chain issue; however, these causes form important starting points for pinpointing changes or stability in demand in future investigations.

\section{Methods and Results}\label{methods}
Mathematical models allow us to gain insight into the dynamics of forces on an outcome variable, in this case house price, that are difficult to measure using historical data. This is because historical data can only represent one state, or realisation, of a system whilst a model can be used to explore all possible states and incorporate drivers that are not directly measurable (e.g. can only be inferred) in a population, for example, demand. Here, we introduce a mathematical model used to gain insight into the dynamics of house price in Australia. In particular, we explore the dynamical relationship between mortgage rate and house price whilst incorporating natural growth (inflation), supply and demand dynamics. We check our model results against historical data on house price (measured using historical mean dwelling value), mortgage rate (measured directly using historical mortgage rate), and supply dynamics (measured using historical dwelling completions). Our model recovers these historical relationships for certain parameter choices and allows us to better understand how a dynamical change in coupling strength of mortgage rate and natural growth can have drastic effects on the house price. Our findings from modelling suggest that a decoupling of mortgage rate occurred due to the low rates set in 2020 and stifling supply. Thus, changes in mortgage rate will likely have little-to-no effect on house price until supply is recovered.

\subsection{A Mathematical Model of House Price}\label{model}

With the insight we gained from our data-driven investigation in Section \ref{data}, we seek to build a mathematical model capable of integrating these relationships into a single system that couples the dynamics of supply, demand, mortgage rate and natural growth to project expectations of house price. Our main differential equation represents the change in house price $Y$ in time according to changes in mortgage rate $X$ and natural linear growth with factor $k$ due to inflation. Recalling that mortgage rate $X$ and house price $Y$ are expected to have a generalised logistic relationship $Y(t) = G(X(t))$ according to historical data-based relationships observed prior to 2020 gives,
\[
\frac{dY}{dt} = \alpha \frac{dG}{dX}\frac{dX}{dt}+(1-\alpha) k
\]
where the strength of the relationships of mortgage rate and linear growth against house price are represented by a coupling parameter $\alpha$. Next, we allow the coupling parameter $\alpha$ to change according to housing supply $S$ and demand $D$ by the smooth Hill-type transition function,
\[
\alpha(S,D) = \frac{(S/D)^2}{(S/D)^2+C^2}.
\]
\begin{wrapfigure}{l}{0.4\textwidth}
\centering
\includegraphics[width=\linewidth]{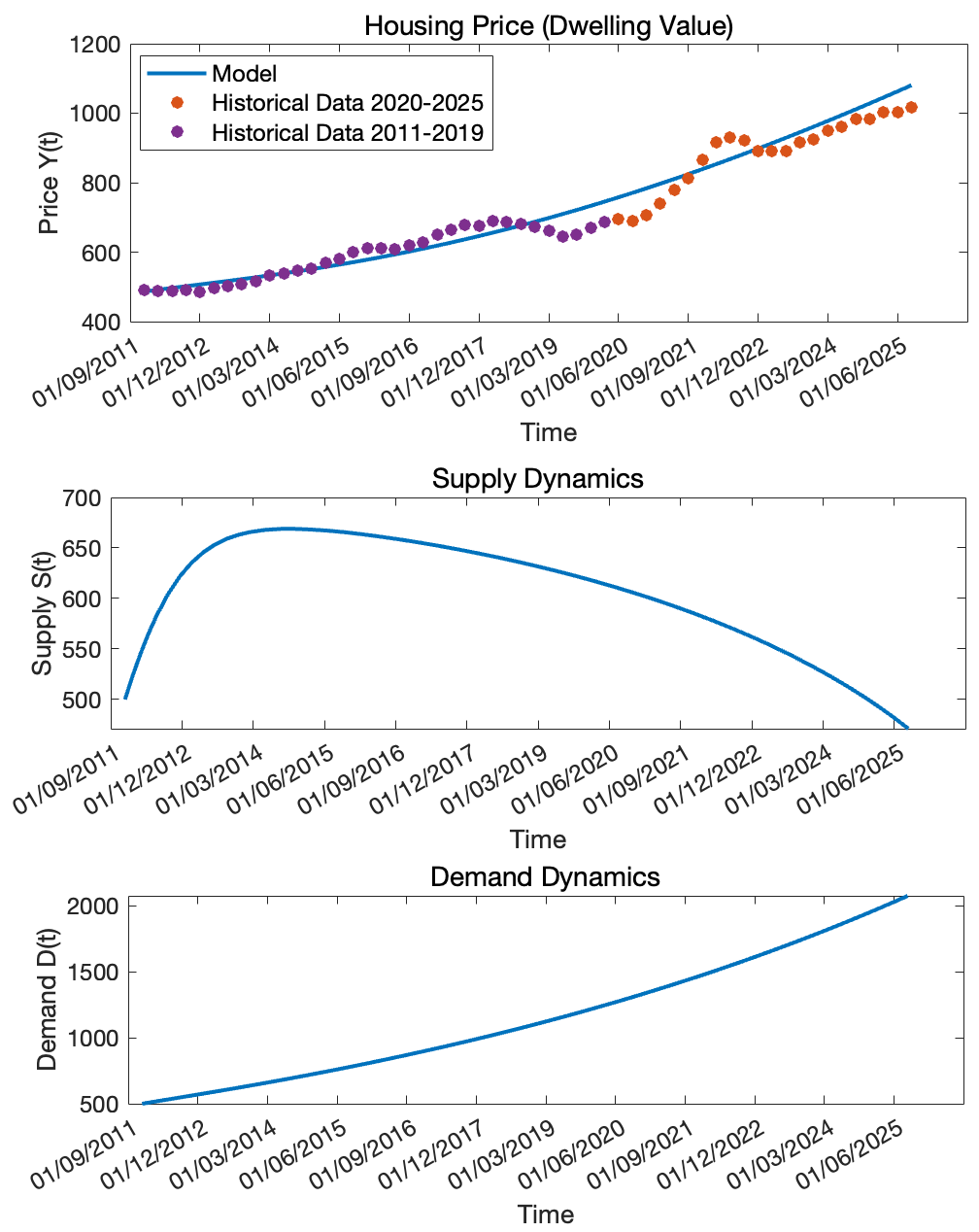}
\caption{Model projections of house price, supply, and demand over time (blue lines). Historical house price (dwelling value) is represented by coloured circles with purple indicating the historical data used to fit the model parameters.}\label{fig:model}
\end{wrapfigure}
Choosing the transition function in this way allows equal contribution of both mortgage rate and inflation relationships when $S =C D$. In the last step, we allow supply $S$ and demand $D$ to change in time according to current values of demand, supply, mortgage rate and house price. We first assume that supply follows a standard logistic growth of parameter $r$ and carrying capacity $K$ with a consumption of supply dictated at a rate $c_S$ proportional to the current house price $Y$. This gives the following differential equation for supply
\[
\frac{dS}{dt} = rS(1-S/K)-c_S Y.
\]
Finally, we allow demand to change only according to mortgage rate $X$ and current demand growth
\[
\frac{dD}{dt} = aX+bD.
\]
We fit the parameters of our mathematical model using historical data from January 2011 - December 2019, prior to the low persistent mortgage rates observed in 2020. A summary of model parameters can be found in Table \ref{tab:model}. We then use the model to project expectations of house price $Y$, supply $S$ and demand $D$ over the interval January 2020 - June 2025 (e.g. the end of our available dataset). Our model's projected values of house price closely mimic those observed over the testing interval (Figure \ref{fig:model}).

\begin{wrapfigure}{l}{0.4\textwidth}
\centering
\includegraphics[width=\linewidth]{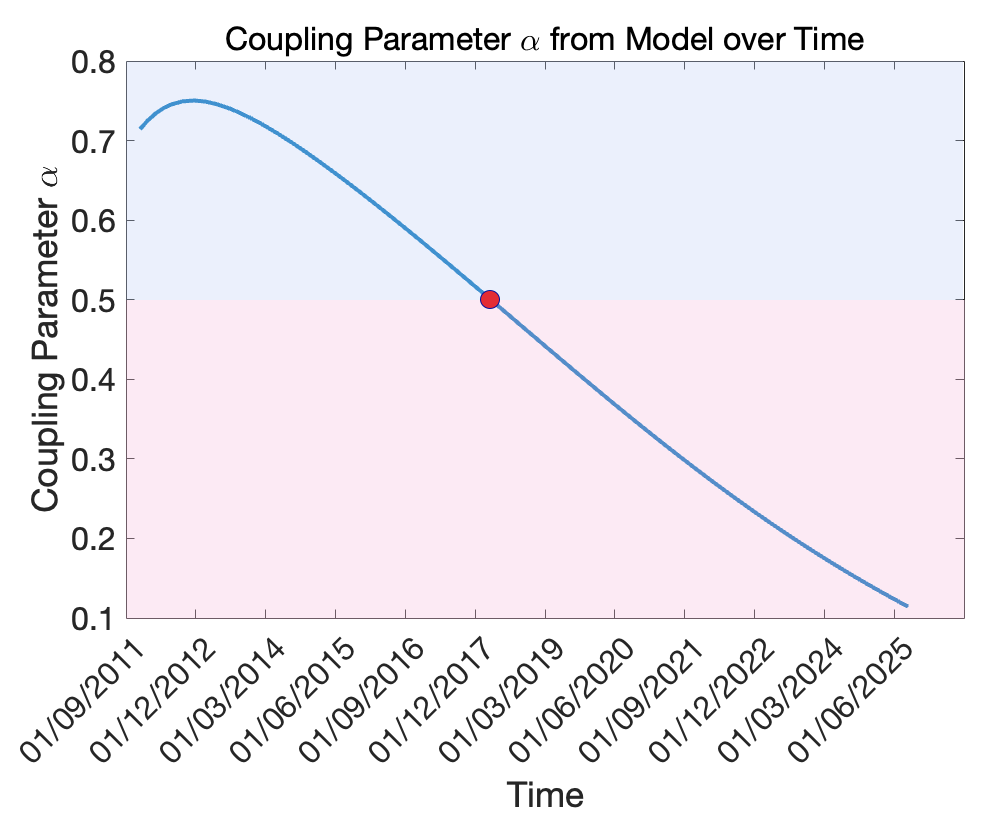}
\caption{Model projections of the coupling parameter $\alpha$ according to model projections of supply and demand.}\label{fig:couple}
\end{wrapfigure}

Several interesting observations emerge from our model. First, the projected supply and demand dynamics indicate a decrease in supply and an increase in demand over time. The coupling parameter $\alpha$ quantifies the relative influence of mortgage rates and inflation on house prices: values $\alpha>>0.5$ indicate that house prices are predominantly driven by mortgage rates, $\alpha = 0.5$ corresponds to equal influence from mortgage rates and inflation, and $\alpha<<0.5$ suggests that inflation is the dominant driver. Using the projected supply and demand dynamics, we observe that $\alpha$ initially increases, reaches a peak, and then declines steadily over time (Figure \ref{fig:couple}). This behaviour suggests a progressive decoupling of the traditional relationship between mortgage rates and house prices. Accordingly, as time increases we observe mortgage rate playing a diminishing role in controlling house price whilst natural linear growth becomes the dominating factor.\\

\begin{table}[h!]
    \centering
    \small
    \begin{tabular}{ccc}
	Parameter & Description & Model Value\\
        \hline
        \hline
        $k$ & linear growth factor of the house price & 20\\
        $G(X)$ & fitted generalised logistic function & $\approx 683.8 -197.7/(1 + 0.5\exp(-5*(X - 5.1)))^{10}$\\
        $C$ & switching parameter for Hill function & $\sqrt{0.4}$\\
        $r$ & logistic growth factor of supply & 0.5\\
        $K$ & carrying capacity of supply & 800\\
        $c$ & consumption of supply & 0.1\\
        $a$ & proportion of demand change in mortgage rate & 1\\
        $b$ & proportion of demand increase in time & 0.02
    \end{tabular}
    \caption{Model parameter values and descriptions. Values were approximated using data prior to 2020.}
    \label{tab:model}
\end{table}
\newpage
\subsection{Forecasting Extreme House Prices}
The objective of this section is to combine the model dynamics observed in Section \ref{model} into a data-driven framework that suitably incorporates real-world uncertainty in the house price. In general, the most appropriate approach is to model the probability distribution of extreme house prices and assess whether the key drivers represented in the model (namely, natural growth through inflation and the mortgage rate) statistically significantly influence the likelihood of observing extreme price levels. Specifically, we aim to examine whether the dependence structure of house price extremes with respect to inflation and mortgage rates differs before and after 2020, when a noticeable decoupling between mortgage rates and house prices emerges in the data (Figure \ref{fig:pricerate}).

Given the limited data available for dwelling value (e.g. house price), related historical time-series data measuring the stock price for the Australian Mortgage Insurance company Helia Group are used as a proxy after confirmation of sufficiently strong correlation ($\rho=0.78$). Historical values for the consumer price index (CPI), as the measure of natural linear growth by inflation, and mortgage rate are used as inputs into our model of the probability distribution of extremes in Helia Group stock price. We split our stock price data into two time periods, 2014-2019 and 2021-2025, and model the probability distribution of extremes in stock price separately for each group. The year 2020 was excluded as a transition year with dynamics too unstable to model using the required deterministic methods.

To model the extremes in stock price, we use a well-studied method from Extreme Value Theory (EVT) called the \textit{block maxima method}. In this approach, we split the data for each group into monthly blocks and keep the maximum over each block to form the frequency distribution. The benefit of this method is that we know that these maxima will always follow a generalised extreme value distribution (GEV), provided the block lengths are long enough to guarantee sufficient decay of correlation between stock price values and the underlying probability distribution is stationary, or does not change across time or another covariate. The second assumption is clearly untrue for our data; however, more advanced methods in EVT allow us to overcome this issue by defining our distributional parameters, the location (mean) $\mu$, scale (variance) $\sigma$, and shape $\xi$, as functions of the covariates, in this case CPI and mortgage rate \cite{Coles}. Statistical methods such as goodness of fit tests, quantile plots and the likelihood ratio test are used as indicators for determining whether the resulting probability distribution accurately models the monthly block maxima of stock price. Finally, these probability distributions are used to forecast how extreme stock price movements are expected to respond under different scenarios of changes in the CPI and mortgage rates.

We fit a nonstationary GEV on monthly maxima of Helia Group stock price for each time period, 2014-2019 and 2021-2025, using the dynamical insights we gained from Section \ref{model} by defining the parameters as relevant functions of the covariates CPI and mortgage rate,
\begin{align*}
\mu = \mu_0 + \mu_1\times\log(\text{Mortgage Rate})+\mu_2\times\text{CPI}\\
\sigma = \sigma_0 + \sigma_1\times\text{CPI}\\
\xi = \xi.
\end{align*}

Our methods yield noticeably different coefficient estimates for the effects of mortgage rate and CPI on the location parameter $\mu$ of the probability distributions of price extremes across both time periods (Table \ref{tab:gev}). To assess the practical implications of these differences, we examine the impact of a 1.0 percentage point increase in the mortgage rate alongside a 10.0 point increase in CPI has on the extreme value distribution for each period.

\begin{figure}
\centering
\begin{subfigure}{0.3\textwidth}
\includegraphics[width=\textwidth]{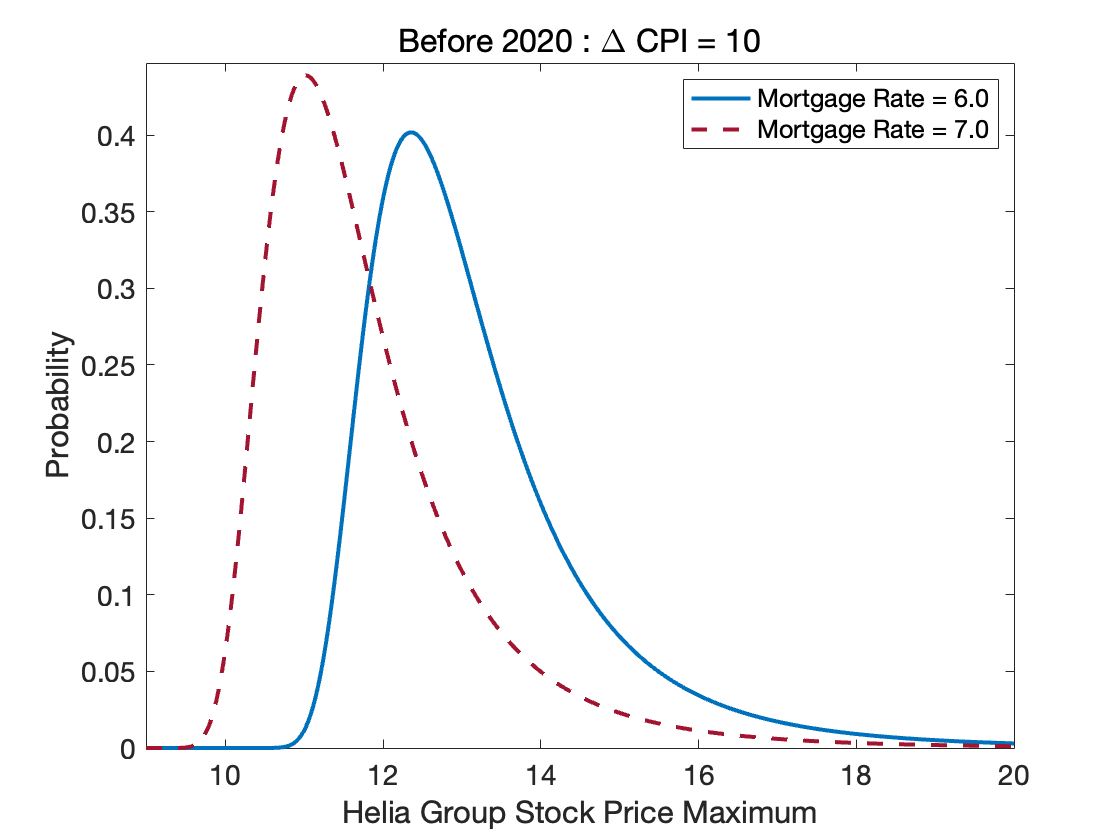}
\caption{}
\end{subfigure}
\begin{subfigure}{0.3\textwidth}
\includegraphics[width=\textwidth]{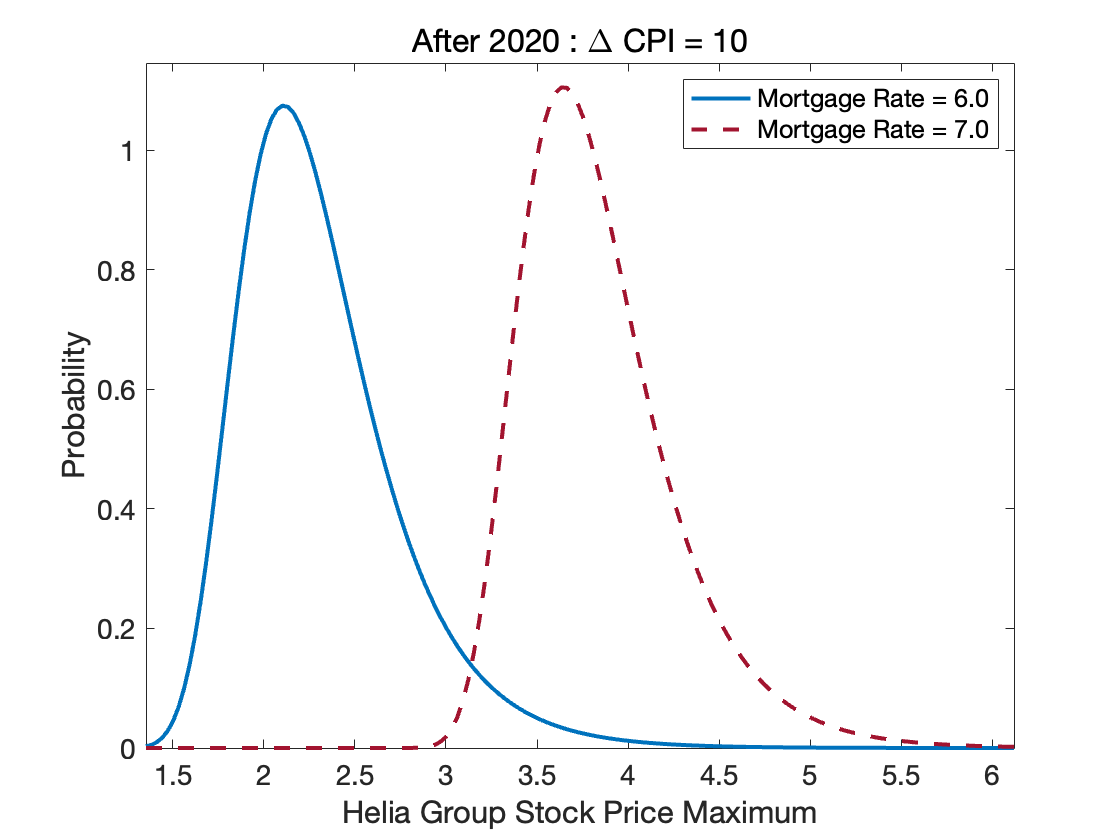}
\caption{}
\end{subfigure}
\begin{subfigure}{0.3\textwidth}
\includegraphics[width=\textwidth]{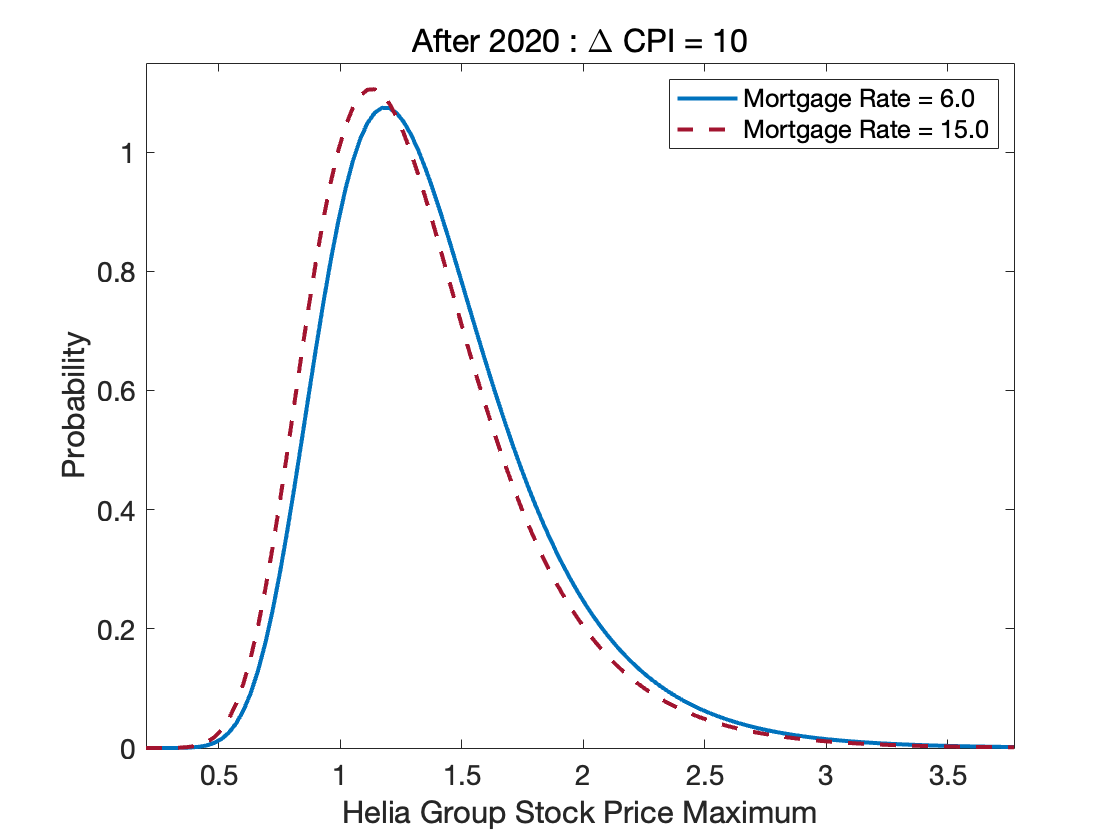}
\caption{}
\end{subfigure}
\caption{Extremal probability distributions for Helia Group price maxima comparing the effect of a 1.0 percentage point mortgage rate increase under a 10.0 point CPI increase for the years (a) 2014-2020 and (b) 2021-2025 and the effect of an 11.0 percentage point mortgage rate increase under a 10.0 point CPI increase for the years (c) 2021-2025.}\label{fig:pd}
\end{figure}

Prior to 2020, a 1.0 percentage point rise in mortgage rate more than offset the effect of a 10.0 point increase in CPI, shifting the distribution of extreme prices to the left and resulting in lower expected extreme price levels. However, this relationship no longer holds after 2020. In the post-2020 period, the same changes produce a rightward shift in the distribution, corresponding to higher extreme price levels (Figure \ref{fig:pd}). Thus, whilst mortgage rates exerted a negative (and economically intuitive) influence on extreme prices before 2020, their moderating effect appears substantially weakened in the later period. This finding is consistent with our broader model results and suggests that mortgage rates may no longer function as a reliable mechanism for controlling extreme price movements.

We further investigated the magnitude of mortgage rate adjustments required to counteract a 10.0 point CPI increase under current market dynamics. Our results indicate that an increase of at least 11.0 percentage points would be necessary to prevent an upward shift in the extreme price distribution (Figure \ref{fig:pd}).
\begin{table}[h!]
    \centering
    \small
    \begin{tabular}{cccccc|cc}
    	\hline
    	2014-2019 & & & & & & &\\
	\hline
	$\mu_0$ & $\mu_1$ & $\mu_2$ & $\sigma_0$ & $\sigma_1$ & $\xi$ & ad-test & ks-test\\
        \hline
        \hline
	$17.76\pm 9.82$ & $-0.54\pm 3.61$ & $-0.13\pm 0.04$ & $1.21\pm 2.24$ & $-0.01\pm 0.02$ & $0.18 \pm 0.22$ & $p = 0.99$ & $p=0.93$\\
	\hline
	2021-2025 & & & & & & &\\
	\hline
	$\mu_0$ & $\mu_1$ & $\mu_2$ & $\sigma_0$ & $\sigma_1$ & $\xi$ & ad-test & ks-test\\
        \hline
        \hline
	$-17.92\pm 3.24$ & $-2.75\pm 0.96$ & $0.20\pm 0.04$ & $0.46\pm 1.20$ & $-0.00\pm 0.01$ & $0.00 \pm 0.21$ & $p = 0.84$ & $p=0.90$\\
	\hline
    \end{tabular}
    \caption{Nonstationary GEV parameter fits for price extremes over the time periods 2014-2019 and 2021-2025.}\label{tab:gev}
    \label{tab:model}
\end{table}

\section{Discussion}\label{discussion}
Our dynamical model provides a framework for understanding how macroeconomic factors and supply constraints interact to drive Australian house prices. Using historical data from January 2011 to December 2019, we fit a model projecting house price, supply and demand through to June 2025. The model reproduces historical trends closely, capturing both supply and demand dynamics over time. A key output of the model is the coupling parameter $\alpha$, which quantifies the relative influence of mortgage rates versus natural linear growth on house prices. Values of 
$\alpha>>0.5$ indicate dominance of mortgage rates, $\alpha\approx 0.5$ indicates equal influence, and $\alpha<<0.5$ suggests that natural growth dominates. Our results show that $\alpha$ initially increases, peaks, and then declines steadily, indicating a progressive decoupling of mortgage rates from house price dynamics. Post-2020, mortgage rates have a substantially weakened effect, while natural linear growth and structural factors drive price trends. 

We extended these insights using a nonstationary generalised extreme value distribution applied to monthly maxima of Helia Group stock price (our proxy for dwelling value) for the time periods 2014-2019 and 2021-2025, with location $\mu$ and scale $\sigma$ parameters defined as functions of mortgage rate and CPI. Prior to 2020, a 1.0 percentage point rise in mortgage rate could offset a 10-point increase in CPI, shifting the distribution of extreme prices to lower values. In contrast, post-2020 projections show that the same changes lead to higher extreme prices, consistent with the decoupling captured by $\alpha$. Our analysis further indicates that preventing upward shifts under current dynamics would require mortgage rate increases of at least 11 percentage points, far beyond feasible policy adjustments.

The weakening of mortgage rate influence is closely linked to structural supply constraints. Our model shows decreasing supply alongside increasing demand over time, highlighting persistent bottlenecks. Supply responsiveness in Australia is influenced not only by private market capacity but also by government actions, including land release, planning approvals, and infrastructure provision. Competition for finite government-managed resources, such as labor, materials, or funding, can slow approvals and construction, further limiting supply. These structural factors reduce the effectiveness of mortgage rate adjustments, as even substantial policy interventions cannot overcome underlying scarcity.

Our results also suggest that demographic factors such as net migration have little measurable effect on dwelling values, reinforcing that national price trends are dominated by supply limitations and macroeconomic conditions rather than population inflows. Together, these findings indicate that restoring any policy leverage over house prices will require targeted efforts to alleviate structural supply constraints and improve coordination in government resource allocation.

\appendix

\begin{table}[h!]
\centering
\small
    \begin{tabular}{p{5cm}p{3cm}p{3.5cm}p{3cm}p{2.5cm}}
	Name & Description & Origin & Available Dates & Frequency \\
        \hline
        \hline
        	Mean Dwelling Value & Mean dwelling value across all dwellings nationally in Australia measured in \$XXX,000 & Australian Bureau of Statistics & January 2011 - June 2025 & Quarterly\\
	\hline
	Dwelling Completion Value & Total value of dwelling completions nationally in Australia measures in \$XXX,000 & Australian Bureau of Statistics & March 1955 - June 2025 & Quarterly\\
	\hline
	Net Migration & Total net migration: sum of border crossings into Australia minus border crossings out of Australia, normalised for transit & Australian Bureau of Statistics & January 1991 - September 2025 & Monthly\\
	\hline
	Mortgage Insurance Stock Helia Group & Stock price measured in \$XX.00 & Bloomberg & January 2014 - December 2025 & Daily
    \end{tabular}
        \caption{Historical data summary table.}\label{tab:ts}

\end{table}
\subsection*{Disclosure Statement} The views, analyses, and conclusions expressed in this investigation are solely those of the authors and do not reflect the views or positions of Barings.
\subsection*{Perceived COI Statement} The authors declare a perceived conflict of interest arising from Ashley Burtenshaw’s employment with Barings. Notwithstanding this affiliation, Barings has no financial or non-financial interest in the design, methodology, analysis, or conclusions of this research.

\clearpage
\bibliography{references}

\end{document}